\documentclass[pra,twocolumn,notitlepage,superscriptaddress,nofootinbib,longbibliography]{revtex4}
        
\usepackage{graphicx}
\usepackage{natbib}
\usepackage{epstopdf}
\usepackage{amsmath}
\usepackage{amssymb}
\usepackage{mathbbol}
\usepackage{xcolor}
\usepackage[colorlinks,citecolor=blue,urlcolor=blue]{hyperref}

\newcommand{\beq}{\begin{equation}}
\newcommand{\eeq}{\end{equation} \smallskip}
\newcommand{\beqy}{\begin{eqnarray}}
\newcommand{\eeqy}{\end{eqnarray} \smallskip}

\newcommand{\bit}{\begin{itemize}}
\newcommand{\eit}{\end{itemize}}
\newcommand{\bmat}{\begin{pmatrix}}
\newcommand{\emat}{\end{pmatrix}}

\newcommand{\ve}[1]{\textbf{#1}}

\begin{document}

\title{Vortex dynamics in a compact Kardar-Parisi-Zhang  system}

\author{A. Zamora}

\affiliation{Department of Physics and Astronomy, University College London,
Gower Street, London, WC1E 6BT, United Kingdom}

\author{N. Lad}

\affiliation{Department of Physics and Astronomy, University College London,
Gower Street, London, WC1E 6BT, United Kingdom}

\author{M. H. Szymanska}

\affiliation{Department of Physics and Astronomy, University College London,
Gower Street, London, WC1E 6BT, United Kingdom}

\email{alex.zamora.soto@gmail.com}

\date{\today}
 

\begin{abstract}
We study the dynamics of vortices in a two-dimensional, non-equilibrium system, described by the compact Kardar-Parisi-Zhang equation, after a sudden quench across the critical region. Our exact numerical solution of the phase-ordering kinetics shows that the unique interplay between non-equilibrium and the variable degree of spatial anisotropy leads to different critical regimes. We provide an analytical expression for the vortex evolution, based on scaling arguments, which is in agreement with the numerical results, and confirms the form of the interaction potential between vortices in this system. 
\end{abstract}

\maketitle


Topological defects play an important role in two-dimensional (2D) critical systems with either U(1) or SO(2) symmetry \cite{altland2010condensed}. A paradigmatic example is the Berezinskii-Kosterlitz-Thouless (BKT) phase transition between disordered and ordered phases of the equilibrium planar XY model, caused by vortices binding at low temperatures due to their mutual attractive interactions \cite{kosterlitz1973ordering,kosterlitz1974critical}. 
Topological defects emerge naturally in a large number of {\it non-equilibrium} systems~\cite{kawasaki1984dynamical,sieberer2018topological,reichhardt2016depinning,thampi2014instabilities}, although their roles at criticality are still largely unexplored.
One paradigmatic case is the compact Kardar-Parisi-Zhang (cKPZ) equation~\cite{sieberer2016lattice,wachtel2016electrodynamic}, which appears as a natural extension of the non-compact KPZ equation, \cite{PhysRevLett.56.889,corwin2012kardar}, when considering  the compactness of the phase.
The cKPZ equation has a wide range of physical applications: from driven-dissipative condensates, such as microcavity polaritons~\cite{PhysRevX.5.011017,PhysRevX.7.041006,he2015scaling,sieberer2016keldysh,sieberer2015thermodynamic};  polar active smectic phases \cite{PhysRevLett.111.088701}; transport phenomena in  periodic  media  (driven vortex  lattices in disordered superconductors  \cite{aranson1998nonequilibrium,balents1998nonequilibrium}); synchronisation  and frequency stability in networks and lattices of coupled limit-cycle oscillators to coupled optomechanical oscillators~\cite{lauter2017kardar}.

Recent theoretical studies suggest that the critical properties of 2D systems governed by the cKPZ equation differ substantially from their equilibrium counterparts, and that the BKT theory  can not be extended to all regimes ~\cite{wachtel2016electrodynamic,Diehl2017private,sieberer2018topological}.
This is rooted in the fact that the vortex-antivortex (V-AV) interactions in an isotropic or weakly anisotropic (WA) cKPZ system, in contrast to the equilibrium XY model, become repulsive beyond a characteristic length scale. In this scenario, a disordered vortex-dominated phase composed of unbound vortices  emerges and precludes a phase transition to a quasi-ordered state in the thermodynamic limit. However, in the strongly anisotropic (SA) cKPZ system, the V-AV interaction is even more attractive than in the XY model, and consequently, the bound vortex pairs stabilise a quasi-ordered phase at low noise levels, to some extent as in the BKT theory for equilibrium systems.

%
%
\begin{figure}
\includegraphics[width=0.9\linewidth]{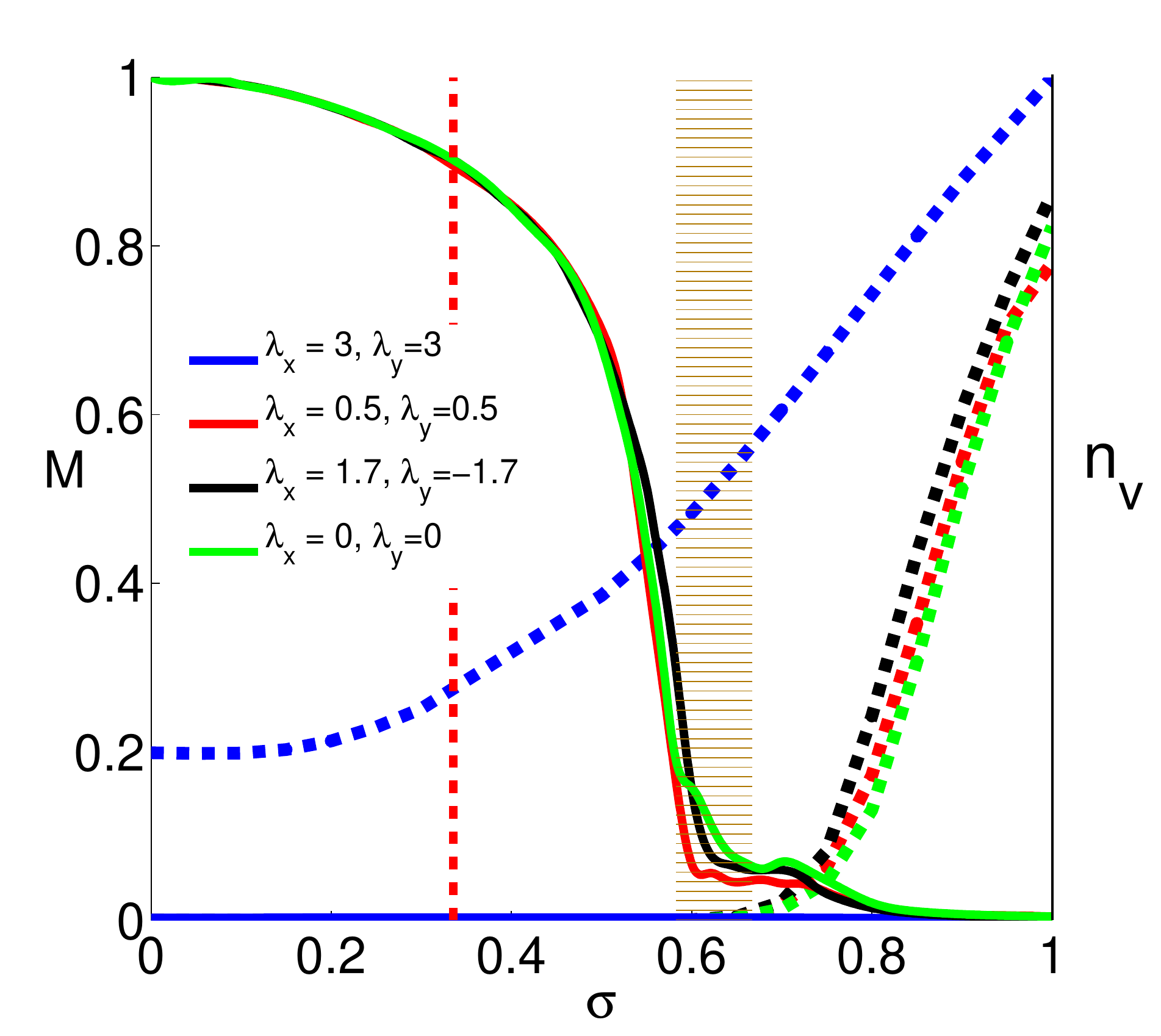}
\caption{\textbf{Steady-state  of the cKPZ system.}
Steady-state density of vortices $\rho$ (in arbitrary units, dashed curves) and magnetization $M$ (solid curves) as a function of the noise parameter $\sigma$ for different $\lambda$. The vertical dashed line indicates the final noise $\sigma_f=1/3$ for the infinite rapid quench shown in Figs.~\ref{fig:sa_regime} and~\ref{fig:wa_regime}. 
For the XY model with $\lambda=0$ (green curves) and the SA case with $\lambda_x=-\lambda_y$ (black curves) the system exhibits (quasi)ordered and disordered phases separated by a critical noise $\sigma_c\approx0.63$ (striped rectangle). However, for the isotropic case with large non-linearity $\lambda$ (blue curves), the system shows a vortex-dominated phase with no magnetisation even at low noise values.
}
\label{fig:st_nv}
\end{figure}
%

In this paper, we present the first clear confirmation of these predictions, based on an exact numerical solution of the full cKPZ dynamics, following a rapid quench across the critical region. We explore the regime of parameters  accessible to the mentioned analytical methods and beyond.  We observe both i) the new  vortex-dominated disordered phase in the WA scenario and ii) the quasi-ordered phase with diminishing number of V-AV pairs in the SA case. 
Finally, we complement our numerical study with an analytical derivation of the vortex equations of motion by considering scaling arguments and the approximate V-AV interaction potential, which is in excellent agreement with the numerics of the full cKPZ dynamics. This confirms that the approximate form of the vortex interactions  captures the essential physics.

\paragraph*{The cKPZ equation  and the vortex interaction}
\label{sec:system}
%
%

The cKPZ equation for compact variable $\theta(\ve{r},t)$ reads~\cite{sieberer2016lattice,sieberer2018topological}
\begin{equation}
\partial_t\theta(\ve{r},t) = \sum_{\substack{ i=x,y}} 
   \Bigl[ D_i \partial_i^2 \theta(\ve{r},t)   +
       \frac{\lambda_i}{2} \Bigl(\partial_i \theta(\ve{r},t) \Bigl)^2
  + \eta(\ve{r},t)  \Bigl],
\label{eq:cKPZ}
\end{equation}
where, $\theta$ may denote \emph{the phase} of the condensate, of the charge-density wave order parameter, the displacement field in a polar active smectic system, the phase field of coupled limit-cycle oscillators.
The diffusion constants $D_x$ and $D_y$ are positive and here taken to be 1, which can be  obtained by an anisotropic rescaling of the lengths. The non-linear parameters $\lambda$ can be either positive or negative and capture the non-equilibrium nature of the system.
The Gaussian noise term with zero mean fulfils $\langle \eta(\ve{r},t)\eta(\ve{r}',t') \rangle = 2\sigma^2 \delta_{\ve{r},\ve{r}'}\delta_{t,t'}$.
Vortices in the system emerge as a consequence of the compactness of the $\theta$ variable in \eqref{eq:cKPZ}, since $\oint \nabla\theta d\ve{l}  = 2\pi n(\ve{r},t)$, with $n(\ve{r},t) \in \mathbb{Z}$~\cite{sieberer2018topological}, where $\ve{l}$ is the contour. Consequently, $n(\ve{r},t)\neq 0$ denotes a vortex with charge $n$, at site $\ve{r}$ and time $t$.

\begin{figure}
\includegraphics[width=0.48\linewidth]
{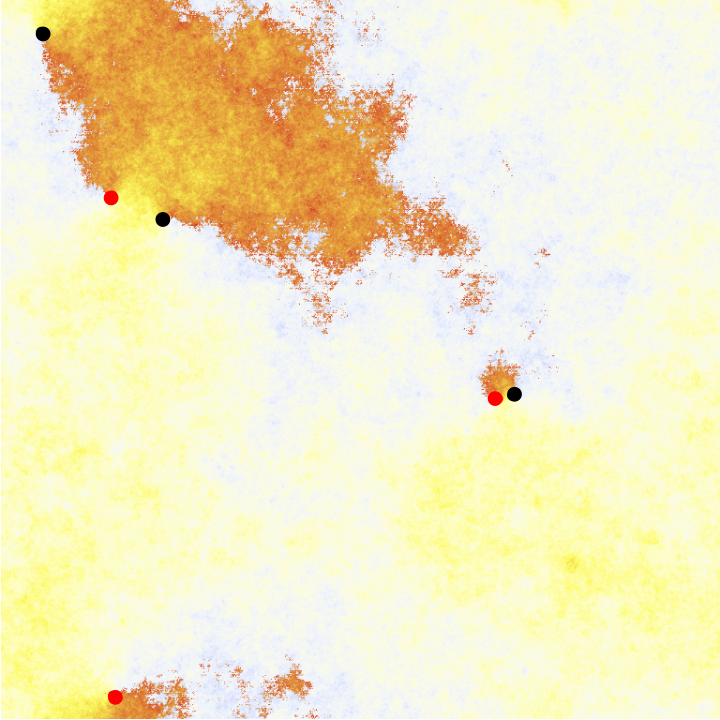}
\includegraphics[width=0.48\linewidth]{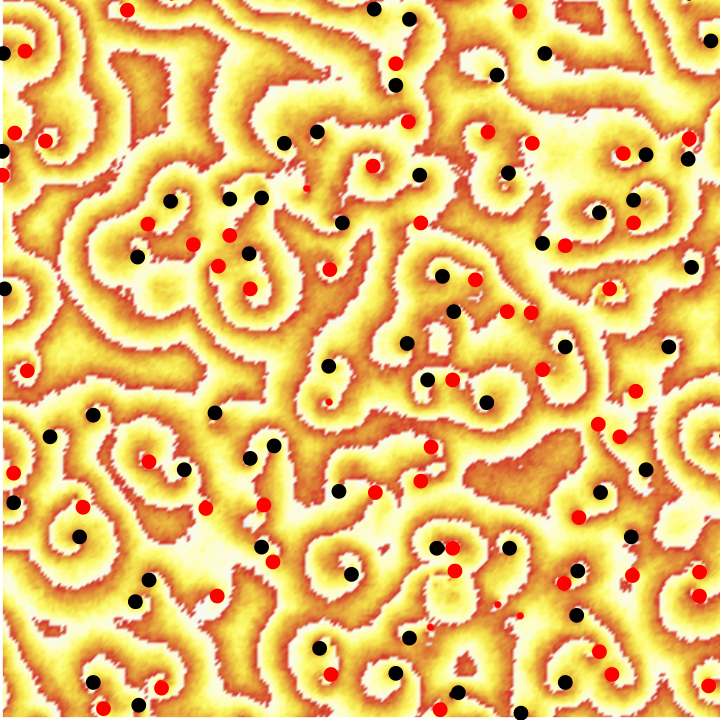}
\caption{\textbf{$\theta(\ve{r})$ at late times with marked vortices.} 2D maps with position of vortices (black dots) and antivortices (red dots) on top of the $\theta(\ve{r})$  profile for a single realisation in two different regimes: i) phase ordering configuration in the SA regime with $\lambda_x=-\lambda_y=1.7$ (\emph{left panel}), note that due to periodic boundaries the red V at the bottom is paired with the black AV on the top, and ii) vortex-dominated phase for $\lambda_x=\lambda_y=1.5$ in the isotropic regime (\emph{right panel}). Note that the left panel shows the whole system, whereas the right panel shows a zoomed quarter  (see the plot for the whole system in Fig.~8 of the Supplementary Material).
}
\label{fig:vortex_and_phase}
\end{figure}    

As has been shown by Sieberer and co-workers~\cite{sieberer2016lattice,wachtel2016electrodynamic,sieberer2018topological}, by considering the dual electrodynamical (dED) picture of the cKPZ equation and a perturbative expansion in the non-linear parameters $\lambda$, the vortices in the cKPZ system interact through a force with both conservative and non-conservative contributions due to the non-equilibrium nature of the system; in contrast to the equilibrium XY model, i.e. when $\lambda_x=\lambda_y=0$, where the vortices interact through only central Coulomb forces ~\cite{minnhagen1987two,ambegaokar1978dissipation}.
However, in the present study we are interested in the scaling and critical properties of the system and, consequently, in the interdistance $R$ between a vortex and an antivortex in a pair. Thus, we consider only the central force within the vortex pair, which can be obtained from the V-AV potential $V(R)$ through $F_{va}(\ve{R}) = -\nabla V(R)$.
This potential, for charge $\pm 1$ vortices reads \cite{sieberer2018topological}:
\beq
V(R) = \frac1\epsilon\log\left( \frac{R}{D_c} \right) -\frac{a}{{3\epsilon^3}}\log^3\left( \frac{R}{D_c} \right),
\label{eq:pot}
\eeq
where $\epsilon$ is the dielectric constant of the non-linear dED theory, $D_c$ is the size of the vortex core and   $a\equiv 2\alpha^2_{+} - \frac{\alpha^2_{-}}{2} + \frac{\alpha_{+}\alpha_{-}}{2}\cos(2\tilde{\theta}) $, with $\tilde{\theta}$ being the angle of the vortex-antivortex dipole, which is set to an average  value of zero in the present study. The $\alpha$ coefficients are  $\alpha_{\pm} = \lambda_{\pm}/(2D)$, with $\lambda_{\pm} = ( \lambda_x \pm \lambda_y )/2$ and $D=D_x=D_y$. 
The first term of the potential \eqref{eq:pot} (zero order in $\lambda$'s, i.e. $\lambda_+ = \lambda_{-}=0$) coincides with the potential of the V-AV interaction in the planar XY model. The second term of \eqref{eq:pot} comes from  the second order correction in the expansion in $\lambda$'s. 
The first order correction in $\lambda$ does not appear in \eqref{eq:pot} since it does not give a central contribution to the force \cite{wachtel2016electrodynamic,sieberer2018topological}.

Differently  to the equilibrium case, where the V-AV interaction  is always attractive, the potential \eqref{eq:pot} can give both attractive and repulsive contributions, depending on the relative sign of the non-linearities $\lambda$ \cite{wachtel2016electrodynamic,sieberer2018topological}. 
Specifically, when both $\lambda$'s have different signs, which defines the SA regime, the force between V and AV is always attractive and enhanced with respect to the analogous force in the equilibrium  XY model.
When the $\lambda$'s have the same sign, which identifies the WA regime, the V-AV force is attractive only up to a given length scale $L_v$ (obtained from the condition $F_{va}(\ve{R})=0$) beyond which it becomes repulsive \cite{wachtel2016electrodynamic,Diehl2017private}. For the isotropic case, where $\lambda_x=\lambda_y=\lambda$
\beq
L_v=D_c \exp\left(2\sqrt{\epsilon}D/\lambda\right).
\label{eq:Lv}
\eeq 
Consequently, it has been predicted that the steady-state of the system shows a vortex dominated phase, characterised by a non-zero density of repelling vortices with a mean interdistance $L_v$~\cite{wachtel2016electrodynamic,Diehl2017private,gladilin2018multivortex}, also observed in the context of the complex Ginzburg-Landau equation \cite{aranson1998spiral}.

\begin{figure}
\includegraphics[width=0.9\linewidth]{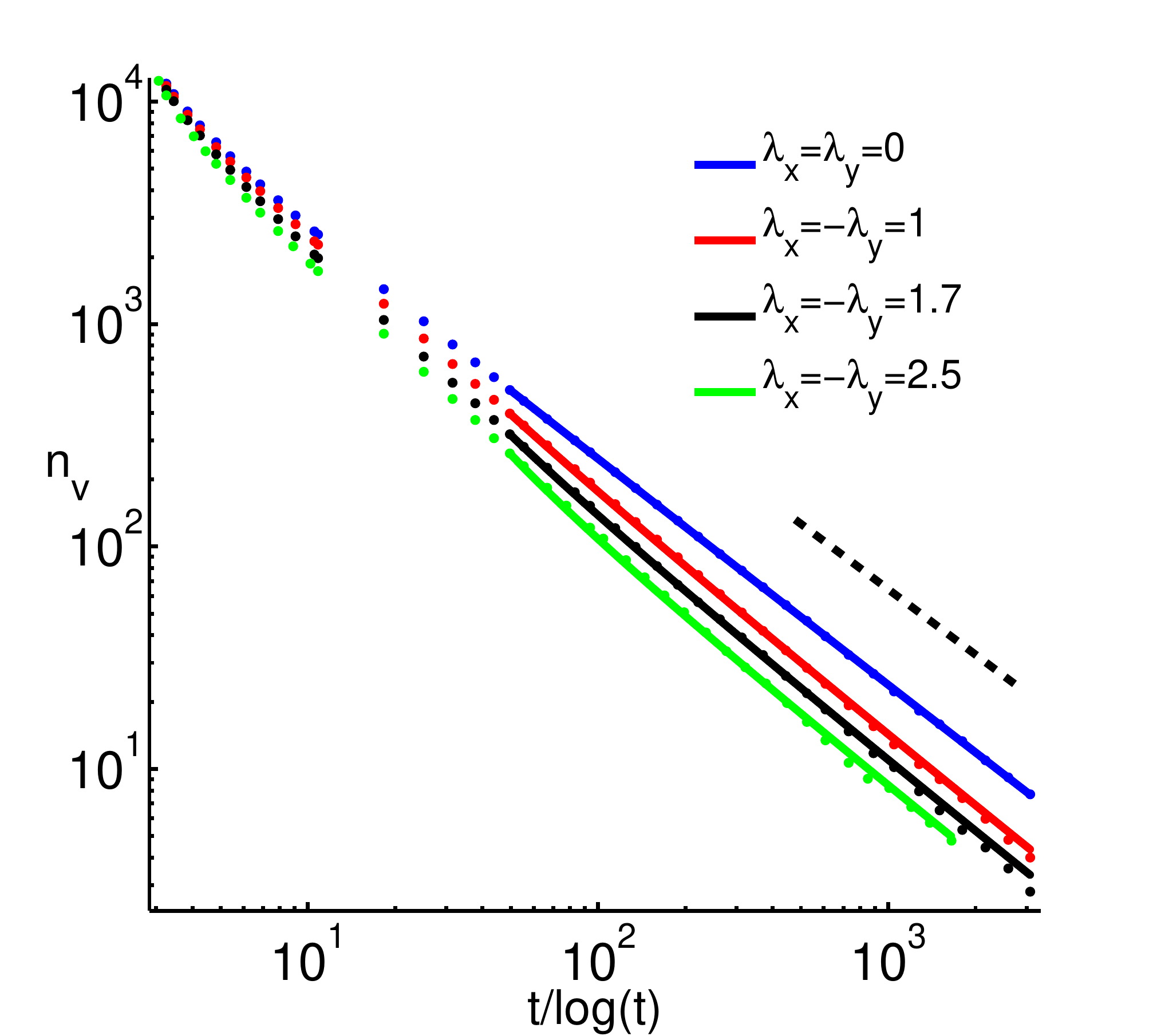}
\caption{\textbf{Vortex dynamics in the SA regime.}
Number of vortices $n_v$ as function of time $t/\log(t)$ after a sudden quench from a high noise disordered initial state to a low noise $\sigma_f$ final state. The dots show the numerical solution of the KPZ equation and the solid lines the theoretical prediction from ~\eqref{eq:xi_time}. We observe a diffusive law with a logarithmic  correction: $n_v\sim (\log(t)/t)^{\alpha}$, with $\alpha\approx 1$ for all the different configurations. Specifically $\alpha=1.016\pm 0.010,\, 1.111\pm 0.011,\, 1.130\pm 0.015,\, 1.142\pm 0.015$ from top to bottom (see Supplementary material for more details). The dashed line shows the $\alpha=1$ case.
}
\label{fig:sa_regime}
\end{figure}

\paragraph*{Dynamical equation for the vortex density}
Considering the vortex potential~\eqref{eq:pot} and general scaling arguments we derive a dynamical equation for the vortex density $\rho$ in the long range limit, which reproduces the numerical integration of the cKPZ equation~\eqref{eq:cKPZ}, revealing that the dynamics of the vortices is governed by the conservative and central forces given by the potential \eqref{eq:pot}.
The starting point is to consider the dynamical equation for a single V-AV pair in the cKPZ system. Assuming that the central potential~\eqref{eq:pot} leads to viscous-relaxation dynamics of the phase (where the forces coming from the central potential are compensated by friction forces on each vortex~\cite{yurke1993coarsening,bray2000breakdown})., for a V-AV pair: $F_{va}+2F_\mu=0$, where $F_\mu=\mu v$ is the friction acting on a single moving vortex with velocity $v$ and inverse mobility $\mu$. Since $v=dD/dt$, where $D$ is the distance between the vortex and antivortex,
leads to \cite{yurke1993coarsening,bray2000random,wachtel2016electrodynamic,reichhardt2016depinning}:
\beq
2F_{\mu}=2\mu(D)\frac{dD}{dt} = -F_{va}(D).
\label{eq:force}
\eeq
The crucial point for characterising the vortex dynamics is the form of $\mu$ since, as in the case of the XY model, we expect a non-trivial dependence on $D$. In this Letter we calculate  $\mu$ associated with a single moving vortex with velocity $\ve{v}$ of modulus $v$, whose field configuration is given by $\phi_\textrm{v}$.
The frictional force reads as $F_{\mu}= -\nabla_{\ve{v}}(dE/dt)$,
with $dE/dt= \int d^2\ve{r}(\delta H/\delta \theta)(d\theta/dt)=\int d^2\ve{r}(d\theta/dt)^2=v^2\int d^2\ve{r} \left(\frac{\partial\phi_\textrm{v}}{\partial_x}\right)^2$~\cite{bray2000breakdown,bray2000random,pleiner1988dynamics}, where $E = \int d^2\ve{r} H$ is the total energy of the vortex configuration with energy density $H$.
Consequently, $\mu = F_{\mu}/v \propto\int d\ve{r} \left(\frac{\partial\phi_\textrm{v}}{\partial_x}\right)^2 \approx \int d\ve{r} \left|\nabla\theta_\textrm{0}\right|^2$, for an isotropic vortex located at the origin in the zero-velocity limit, whose field is given by $\theta_0$.

\begin{figure}
\includegraphics[width=1.0\linewidth]{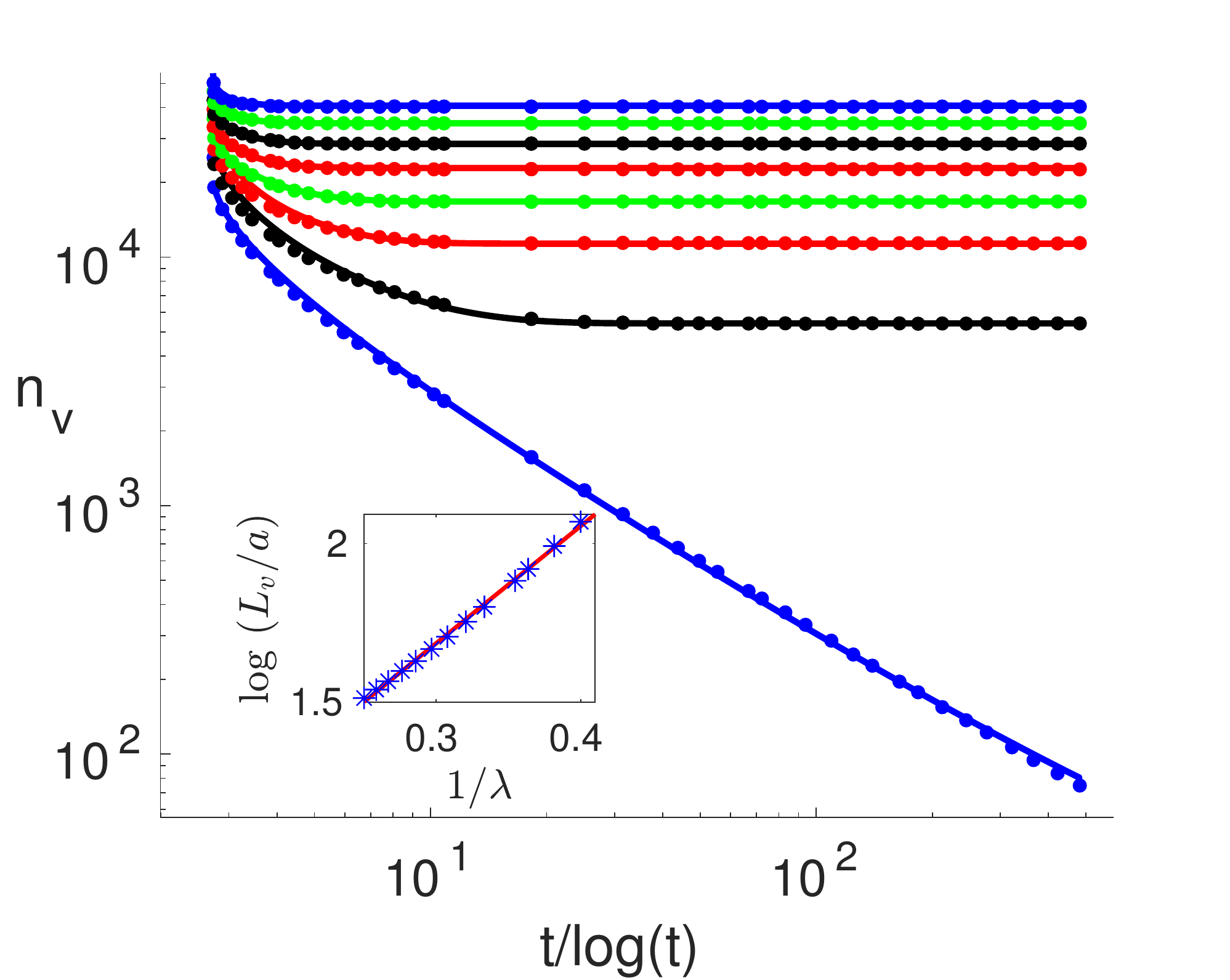}
\caption{\textbf{Vortex dynamics in the WA regime.}
Number of vortices $n_v$ as a function of time $t/\log(t)$ after a sudden quench from a disordered initial state to a low noise $\sigma_f$ final state. The dots are from the numerical solution of the cKPZ equation and the solid lines are the analytical estimate from  \eqref{eq:xi_time}. From top to bottom: $\lambda=3.5,3.25,3.0,2.75,2.5,2.25,1.9,0.5$, with $\lambda\equiv\lambda_x=\lambda_y$. 
We observe a saturation of the number of vortices at $\rho_S(\lambda)$ for $\lambda\geq1.9$, which is a clear indicator of the vortex-dominated phase (see Supplementary Material). Inset: Blue dots show the characteristic length-scale  $L_v\sim \rho_S^{-1/2}$ obtained from numerics and red solid line from the expression \eqref{eq:Lv} for $2.5 \leq \lambda \leq 4.0$.
}
\label{fig:wa_regime}
\end{figure}
%
%
%
Finally, we calculate $\nabla\theta_\textrm{0}$ with the help of the dED relation $\nabla\theta_\textrm{0}=z\times\ve{E}_\textrm{0}$~\cite{wachtel2016electrodynamic,sieberer2016lattice}, where $\ve{E}_\textrm{0}=-\nabla V_v(\ve{r})$ is the electrostatic field created by a single charge (vortex) located at the origin, and the potential $V_v(\ve{r})$ coincides with potential ~\eqref{eq:pot}, since we are considering vortices with charge one. 
Therefore, we find that
%
\begin{multline}
\mu(r)\propto\Bigg\{\log\left(\tilde{r}\right)- 
\frac{2}{3} \tilde{a} \log^3\left(\tilde{r}\right)
+
\frac{1}{5}\tilde{a}^2\log^5\left(\tilde{r}\right)
 \Bigg\},
\label{eq:mu}
\end{multline}
where $\tilde{r}\equiv r/D_c$ and $\tilde{a}=a/\epsilon^2$. Expression~\eqref{eq:mu} contains the characteristic $\log(\tilde{r})$ dependence of the equilibrium XY model ($\lambda_x=\lambda_y=\tilde{a}=0$) followed by higher logarithmic powers from the non-linearity  $\lambda$, characteristic of the KPZ system.
 We now derive the dynamical equation for the vortex density $\rho$ by considering that the system in the long range limit is characterized by a unique single length scale $\xi$, a characteristic velocity $v_\xi=d\xi/dt$, and characteristic elastic $F_{va}(\xi)$ and viscous $\mu(\xi)d\xi/dt$ forces \cite{yurke1993coarsening,bray2000breakdown}.
Taking an absolute value of \eqref{eq:force} we obtain
\beq
\mu(\xi)\frac{d\xi}{dt} \propto \frac{1}{\xi}
\left| 1 - \tilde{a}\log^2\left( \frac{\xi}{D_c} \right) \right|,
\label{eq:xi_time}
\eeq
which leads to the vortex density dynamics in the long range limit through the relation $\rho\sim1/\xi^2$ and using~\eqref{eq:mu}.

\paragraph*{Vortex dynamics after an infinitely rapid quench}

Here,  we study the dynamics of the vortex density after an infinitely rapid quench from a completely disordered phase to a low noise regime with $\sigma_f=1/3$ (see Fig.~\eqref{fig:st_nv}) by exact numerical solution of full cKPZ equation~\eqref{eq:cKPZ}.
Phase ordering kinetics following this type of quench protocol has been used to study universal properties of both equilibrium ~\cite{bray2002theory,yurke1993coarsening,bray2000breakdown,jelic2011quench,godreche2000response} and non-equilibrium~\cite{comaron2017dynamical,kulczykowski2017phase} complex systems.

\subparagraph*{Steady-state }
Before addressing the quench dynamics, we first characterize the non-equilibrium steady state (Fig. \ref{fig:vortex_and_phase}). 
We find that in the SA regime,  characterised by different signs of $\lambda_x$ and $\lambda_y$,  the system shows two distinct  phases in the steady-state: i) A phase with vanishing density of vortices at low noise levels below a critical noise $\sigma_c$ (left panel in Fig. \ref{fig:vortex_and_phase}); and with a finite magnetisation  $M^2(t)= \frac{1}{V_s}(m_x^2 + m_y^2$), where $V_s$ is the volume of the system, $m_x=\langle\int d^2\ve{r}\cos\theta(\ve{r},t)\rangle$, $m_y=\langle\int d^2\ve{r}\sin\theta(\ve{r},t)\rangle$, and $\langle \cdot\cdot\cdot\rangle$ denotes averaging over stochastic realisations. Note that $M$ depends slightly on $\lambda$.  
ii) A disordered phase, characterised by a high density of  vortices, which destroy the magnetisation at noise levels above $\sigma_c$. This critical behaviour shown in Fig.~\ref{fig:st_nv}, equivalent to the one present in the equilibrium XY model ($\lambda_x=\lambda_y=0$), is consistent with analytical predictions by Sieberer et al.  using approximate methods \cite{sieberer2018topological}.

However, this picture changes completely at the WA regime, i.e. $\lambda_x$ and $\lambda_y$ with same sign.   We find that the steady state shows a non-zero density of topological defects, which destroy the magnetisation,  even at low and vanishing noise levels when $\lambda\geq 2.5$, with $\lambda=\lambda_x=\lambda_y$ (see Fig.~\ref{fig:st_nv} and right panel of  Fig. \ref{fig:vortex_and_phase}).
The transition between a (quasi)ordered phase and disordered phases is gone.
We obtain that the density of vortices is independent of the noise at low noise strengths, revealing the length scale expressed in~\eqref{eq:Lv} (see Supplementary material for details), which is a clear indicator of the vortex dominated phase predicted by Sieberer et al.~\cite{sieberer2018topological,wachtel2016electrodynamic}, and analogous to  one identified in the context of the complex Ginzburg-Landau equation \cite{aranson1998spiral}.
We also observe that for low values of $\lambda$, the characteristic length $L_v$ exceeds the system size considered in this study and, consequently, the system does not exhibit the vortex dominated phase. In contrast, it shows two distinct phases as in the SA case, one with a finite $M$ and low vortex density (at low noise levels) and another with a high density of entropic vortices, which destroy the magnetisation of the system (see the $\lambda_x=\lambda_y=0.5$ case in Fig.~\ref{fig:st_nv}). This behaviour is a finite size effect since $L_v>L$ in this case, where $L$ is the system size.

\subparagraph*{Diffusive decay of the vortex density }

Firstly, we consider an infinitely  rapid quench through a critical point in the SA regime for different cases. Our exact numerical solutions of the cKPZ equations  show that the vortex density decays in time following a diffusive law with a logarithmic correction as in the equilibrium XY model, i.e. $\rho \sim (log(t)/t)^{\alpha}$ \cite{bray2000breakdown,jelic2011quench}, as we can see in Fig. \ref{fig:sa_regime}.
(for further details of the fit and discussion we refer to Fig.~7 and Sec. II and IV of Supplementary Material).
An example of late time dynamics  for  $\lambda_x=-\lambda_y=1.7$ case is shown in the left panel of Fig. \ref{fig:vortex_and_phase} for a configuration with 3 pairs of vortices  and in Fig.~8 of the Supplementary Material for a configuration with no vortices). 
There is a very good agreement between the vortex dynamics coming from our numerics and the dynamics predicted by Eq.~\eqref{eq:xi_time}. 
For all values of $\lambda$ considered here,
the exponent $\alpha\approx1$ (the difference between the values of the exponents for the $\lambda\neq 0$ and the $\lambda_x=\lambda_y=0$  is less than $15\%$.
We, however, notice a weak dependence of the exponent $\alpha$ on $\lambda$ with $\alpha$ for $\lambda\neq 0$ being close to $1.1$ (See Supplementary material for more details). It is not clear whether this deviation from  $\alpha=1$  is due to the non-universal corrections, which could be attributed 
to the enhancement of the V-AV interaction when increasing $\lambda$, or whether the system falls into a different universality class with a critical exponent
$\alpha=1.1$ rather then $\alpha=1$. This will require further investigations.

\subparagraph*{Vortex-dominated phase}

The isotropic system shows a completely different behaviour.
Firstly, for low values of $\lambda\equiv\lambda_x=\lambda_y$, the decay of the vortex density scales asymptotically with the inverse  time $\rho \sim 1/t^{\beta}$.
In contrast to the SA case, the exponent $\beta$ can be much smaller than 1 and depends strongly on $\lambda$ (see  for example the $\lambda=0.5$ case in Fig.~\ref{fig:wa_regime}). 
Note, that again the theoretical prediction of the dynamics of the vortices given by Eq.~\eqref{eq:xi_time}  is in very good agreement with numerical solutions of the cKPZ equation, as displayed in Fig.~\ref{fig:wa_regime}.
Specifically, we observe that $\beta$ decreases sharply when increasing $\lambda$, and becomes $\beta\approx 0$ for $\lambda \geq 1.5$, which indicates that the system reaches a steady-state with a non-zero density of vortices $\rho_S$ (see the right panel of Fig \ref{fig:vortex_and_phase} where we can observe the distinctive spiral configuration of the V-AV phase as predicted in \cite{sieberer2018topological}).
We believe that this saturation in the vortex density is a consequence of the repulsive interactions between the Vs and AVs in the isotropic and  WA regime at large distances.
The characteristic length scale we obtain from the numerical simulations through $\rho_S \sim 1/L_v^2$ \cite{Diehl2017private} (the inset panel of Fig. \ref{fig:wa_regime}) agrees extremely well with the theoretical exponential dependence on $1/\lambda$~\cite{sieberer2018topological,wachtel2016electrodynamic} derived in \ref{fig:wa_regime} (see Supplementary material for a technical discussion) and so indicates the emergence of the disordered vortex dominated phase. 
Finally, we should stress that the lack of saturation in the number of vortices for the $\lambda=0.5$ case is a finite size effect i.e. the length scale associated with the inverse of the steady-state vortex density exceeds the size of the system considered in the present work. We expect a saturation of the number of vortices for all values of $\lambda$ for an infinite system.

\paragraph*{Summary and outlook}

We have explored the crucial role of topological defects in the critical behaviour of a non-equilibrium system described by the cKPZ equation by determining numerically the full dynamics after a sudden quench through a critical point. We have also derived an analytical expression for the vortex density dynamics, using the approximate form of the vortex - anti-vortex potential \cite{sieberer2018topological}, which is in excellent agreement with the numerical results.

Crucially, in the isotropic or WA KPZ regime, i.e. when there are non-vanishing non-linear terms of the same sign in both spatial directions, we have identified a phase characterised by a saturation of the vortex density in the phase ordering process.
This novel behaviour, with no counterpart in equilibrium systems, arises in the non-equilibrium scenario due to the external drive and dissipation, and can be strongly modified by spatial anisotropy.
We believe that this vortex dominated phase appears as a consequence of the repulsive V-AV interactions at large distances. Our results confirm the existence of a new vortex-dominated phase in systems larger then a characteristic length scale, which is exponentially dependent on the inverse of the non-linear KPZ parameter.

In the opposite scenario, i.e. with either vanishing non-linear terms or with no-linearities of opposite sign in the two spatial direction (SA regime), we find that the vortex density decays in time algebraically with an exponent close to $-1$ and logarithmic corrections due to the attractive V-AV interactions, as in the equilibrium planar XY model scenario.

Since the cKPZ equation describes the behaviour of a wide range of atomic, molecular and optical systems, it would be of a great interest to obtain the parameters of this equation from microscopic analysis of a particular realisation. This would allow us to determine whether the new vortex dominated phase as well as the transition between isotropic or WA and SA regimes can be practically realised in any of these realistic scenarios.
Finally, the impact of the sign of the V-AV interaction on the vortex dynamics when the system crosses a critical point by following a {\it finite} quench has not yet been explored. This type of critical dynamics, which is successfully described for equilibrium systems by the Kibble-Zurek mechanism~\cite{kibble1976topology,zurek1985cosmological}, can have different properties out of equilibrium, and is not to date clear whether the extension of the concept of adiabaticity could be carried over to the non-equilibrium scenarios~\cite{hedvall2017dynamics}.

\paragraph*{Acknowledgements}

We would like to thank L. Sieberer and S. Diehl for fruitful discussions and A. Ferrier for a revision of the text.  We acknowledge financial support from the EPSRC (Grants No. EP/I028900/2 and No. EP/ K003623/2).


\widetext
\begin{center}
	\textbf{\large Supplementary Material for: Vortex dynamics in a compact Kardar-Parisi-Zhang  system}
\end{center}

\vspace{8mm}
\twocolumngrid

In this supplementary material we discuss in detail different relevant technical aspects concerning the numerical methods that we use in the present study.
In particular, we describe the numerical method for integrating the cKPZ equation and discuss the convergence of the results with respect to the time step, the system size, and the number of stochastic realisations, both for the steady-state and the dynamics after an infinitely rapid quench across the critical point.
%

\section{Compact KPZ equation on a lattice: numerical methods}
\label{sec:convergence}

We consider a system on a two dimensional $N\times N$ lattice, with $N=1024$ and lattice spacing $a=1$. Consequently, we numerically solve the discrete version \cite{sieberer2016lattice} of the continuous  cKPZ equation (1) in the main text:
\begin{equation} 
\partial_t\theta_{\ve{r}} = -\sum_{\substack{ i=x,y \\ \ve{a}_i }} 
   \Bigl[ D_i\sin \Delta\theta_{\ve{a}_i}   + 
       \frac{\lambda_i}{2} ( \cos \Delta\theta_{\ve{a}_i} - 1 )
  + \eta_{\ve{r}}  \Bigl],
\label{eq:cKPZLattice}
\end{equation}
where $\Delta\theta_{\ve{a}_i} = \theta_{\ve{r}}(t) - \theta_{\ve{r}+\ve{a}_i}(t)$ is the difference between the phase at $\ve{r}$ and at the nearest neighbour site at $\ve{r} + \ve{a}_i$, with unit vector $\ve{a}_i$. As in the main text: i) the diffusion constants $D_x$ and $D_y$ are set to 1 ii) the non-linear parameters $\lambda$ can be either positive or negative iii) the Gaussian noise term with zero mean fulfils $\langle \eta_{\ve{r}}(t)\eta_{\ve{r}}(t') \rangle = 2\sigma \delta_{\ve{r},\ve{r}'}\delta_{t,t'}$.

\begin{figure}
\includegraphics[width=0.95\linewidth]{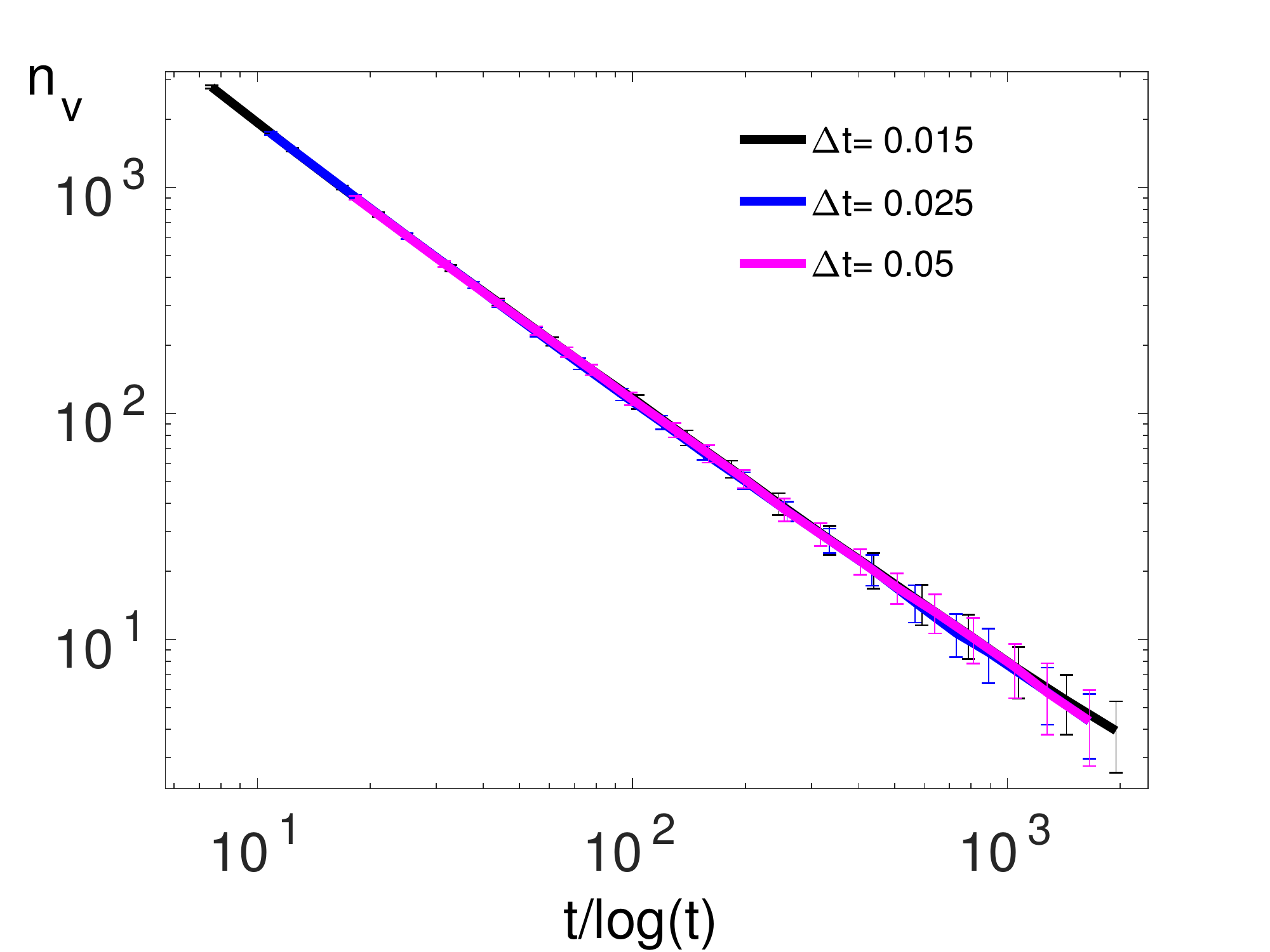}
\caption{\textbf{Phase ordering kinetics for different time steps $\Delta t$.}
Number of vortices as a function of time ($t/\log t$) for the $\lambda_x=-\lambda_y=2.5$ configuration, for different time steps $\Delta t$.
}
\label{fig:xy_decay}
\end{figure}

We use the Euler-Maruyama method for solving \eqref{eq:cKPZLattice} on a lattice~\cite{higham2002strong,greiner1988numerical,moser1991numerical,miranda2008numerical}, which states that the stochastic differential equation expressed as 
$
dX_t = a(X_t) + b(X_t)dW_t, 
$
where $W_t$ represents a Wiener process, can be numerically solved in discrete time, with time step $\Delta t$ and total time iterations $N_{\textrm{time}}$ (where the physical time $t$ is given by $t=\Delta t\cdot N_{\textrm{time}}$), by making use of finite differences: 
$
X_{t+\Delta t} = X_t + a(X_t)\Delta t + b(X_t)\Delta W_t, 
$
where $\Delta W_t=W_{t+\Delta t }-W_t$. Specifically in our case, $\Delta t=0.025$, $X_t=\theta_{\ve{r}}(t), \;
a(X_t)=-\sum_{\substack{ i=x,y \\ \ve{a}_i }} \Bigl[ D_i\sin \Delta\theta_{\ve{a}_i}   + \frac{\lambda_i}{2} ( \cos \Delta\theta_{\ve{a}_i} - 1 )\Bigl]$ 
and $\Delta W_t = \xi\sqrt{12\Delta t}R $, where $\xi^2=2\sigma$ and $R$ is a random number uniformly distributed between $-1/2$ and $1/2$. The prefactor $\sqrt{12\Delta t}$ ensures that the noise has the same second moment as the Gaussian noise over $\Delta t$ \cite{moser1991numerical,miranda2008numerical}.  For the results presented in the main text we have averaged over 200 stochastic realisations.

Note that the compactness of the phase variable $\theta_{\ve{r}}$, which results in the appearance of the vortices, is guaranteed by the action of the \emph{sin} and \emph{cos} trigonometric functions and by the addition of $2\pi m_\ve{r}$ to $\theta_{\ve{r}}(t+\Delta t)$, with $m_\ve{r} \in\mathbb{Z}$, such that $\theta_{\ve{r}}(t+\Delta t)\in [0,2\pi)$ \cite{sieberer2016lattice,wachtel2016electrodynamic}.

\section{Compact KPZ equation on a lattice: convergence with respect to $\Delta t, \; N$ and $M$}
\label{sec:convergence}

In the following, we discuss the convergence of the physical results shown in the main text with respect to different parameters appearing when solving the lattice cKPZ equation~\eqref{eq:cKPZLattice}. Specifically, we consider the convergence with respect to the time step $\Delta t$, the systems size $N$ and the number of stochastic realisations $M$.

In Fig.~\ref{fig:xy_decay} we display the dynamics of the vortex number during the phase ordering process for a SA configuration  with $\lambda_x=-\lambda_y=2.5$ for $\Delta t = 0.015, 0.025, 0.05$.
The  convergence of the vortex number is reached for $\Delta t \leq 0.025$.
We also study the dependence of the  exponent $\alpha$ of the XY model~\cite{yurke1993coarsening,bray2000breakdown,bray2000random}:
\beq
n_v(t)\sim (logt /t)^{\alpha},
\label{eq:xy_decay}
\eeq
on the time step $\Delta t$.
In Table~\ref{tab:xy_exp} we compare the expected theoretical value $\alpha=1$ with the numerical results 
 as a function of the time step  $\Delta t$.
\begin{table}[bt]
\begin{tabular}{c|c|c|c}
$\Delta t$   &   $\alpha$  &  $\pm\Delta\alpha$  &  deviation ($\%$)   \\
\hline\hline
\;\; $0.015$ \;\; &  \;\; $0.97$ \;\;  &  \;\; $0.07$ \;\; &  $3.0$  \\
\hline
    $0.025$  &   $1.02$  &   $0.10$  &  $2.0$  \\
\hline
    $0.050$  &   $1.05$  &   $0.16$  &  $5.0$  \\
\hline
    $0.10$   &   $1.06$  &   $0.23$  &  $6.0$  \\
\hline
    $0.15$   &   $1.05$  &   $0.25$  &  $5.0$  \\
\hline
\end{tabular}
\caption{\textbf{Exponent for the decay of the vortex number in the XY model}. 
The first column corresponds to the time step $\Delta t$ and the other three columns show the exponent $\alpha$, its uncertainty given by the statistical analysis of the data and the deviation with respect to the theoretical value $1$, respectively.
We consider the expression~\eqref{eq:xy_decay}, for times $80<t<34000$. } 
\label{tab:xy_exp}
\end{table}
\begin{figure}
\includegraphics[width=1.0\linewidth]{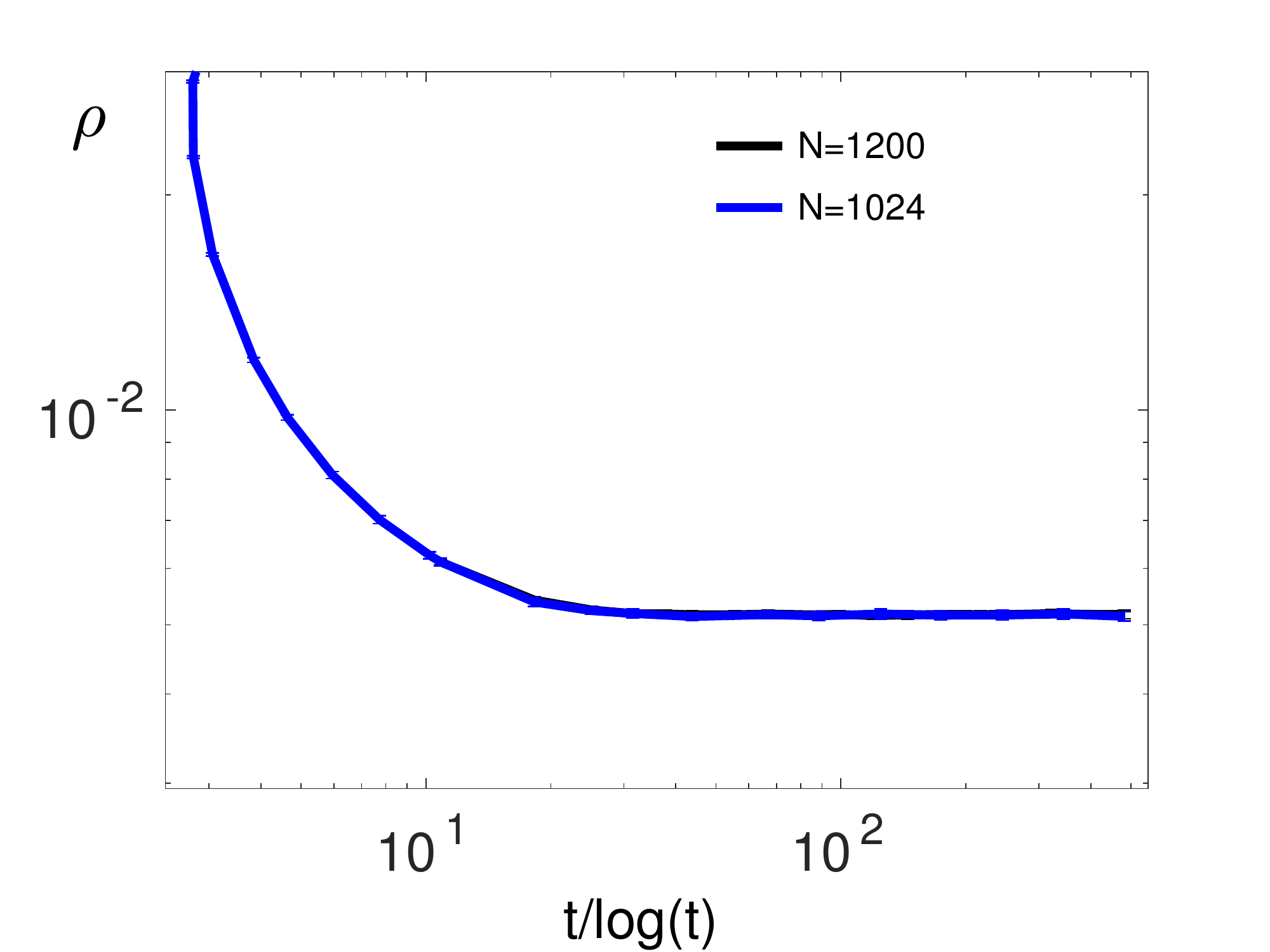}
\caption{\textbf{Vortex density for two system sizes.}
Vortex density $\rho$ as a function of time ($t/\log t$) for $\lambda_x=\lambda_y=1.9$ with $N=1024$ (blue curve) and $N=1200$ (black curve).
}
\label{fig:convergenceLattice}
\end{figure}
%
%
We observe that the deviation $| (\alpha-1)/1 | $ (4th column) of the numerical value for $\alpha$ from the theoretical value $1$ does not decrease for time steps smaller then $\Delta t=0.025$.
%

\begin{figure}[ht]
\includegraphics[width=0.9\linewidth]{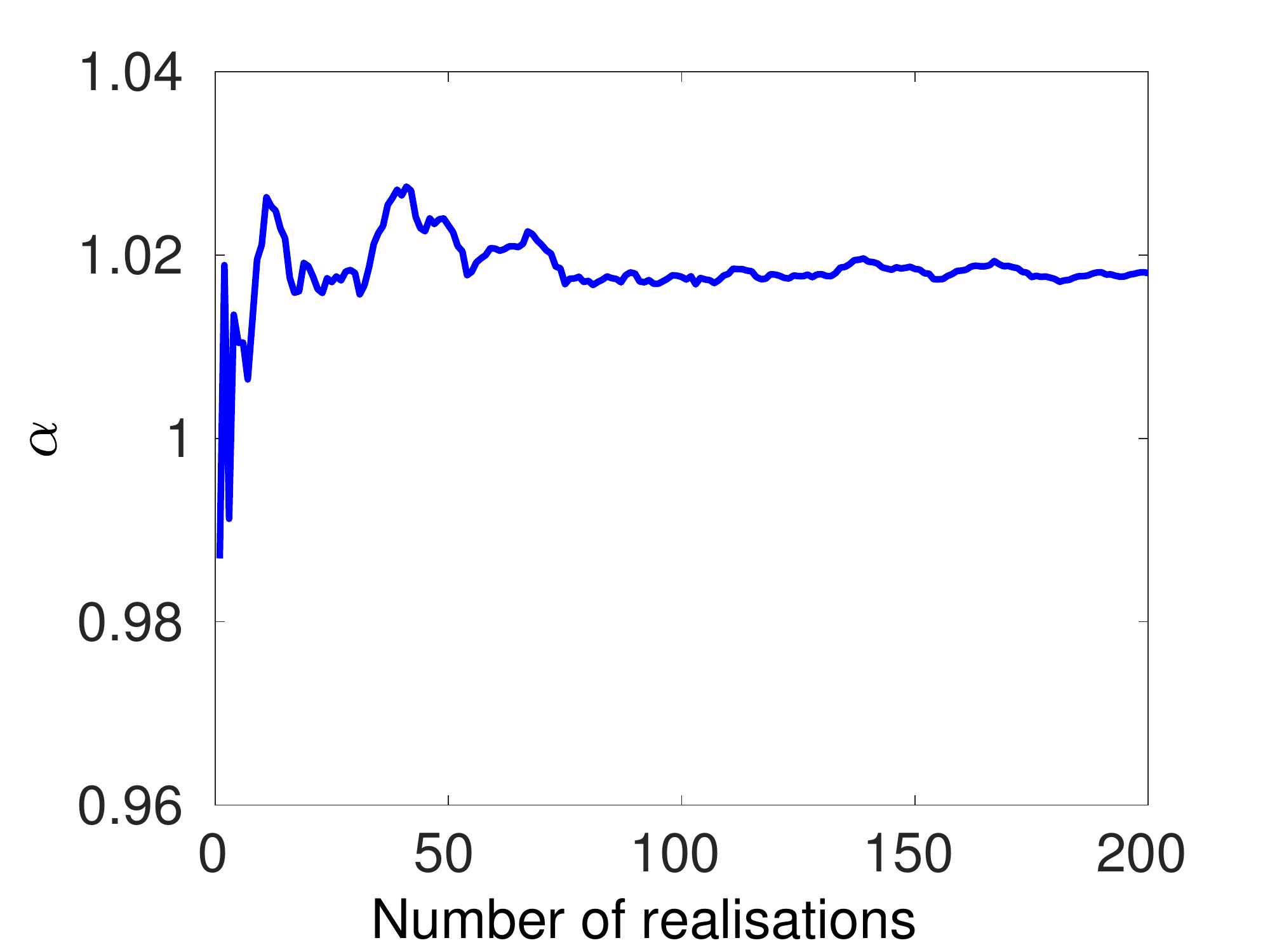}
\caption{\textbf{Convergence of the universal exponent with the number the stochastic realisations.}
Exponent $\alpha$ (see Eq.\eqref{eq:xy_decay}) for the XY model as a function of the number of stochastic realisations $M$ for times within the interval $[80,34000]$. We observe a convergence of the critical exponent for $M>100$.
}
\label{fig:convergenceRealisations}
\end{figure}

%
%
\begin{figure}[ht]
\includegraphics[width=0.9\linewidth]{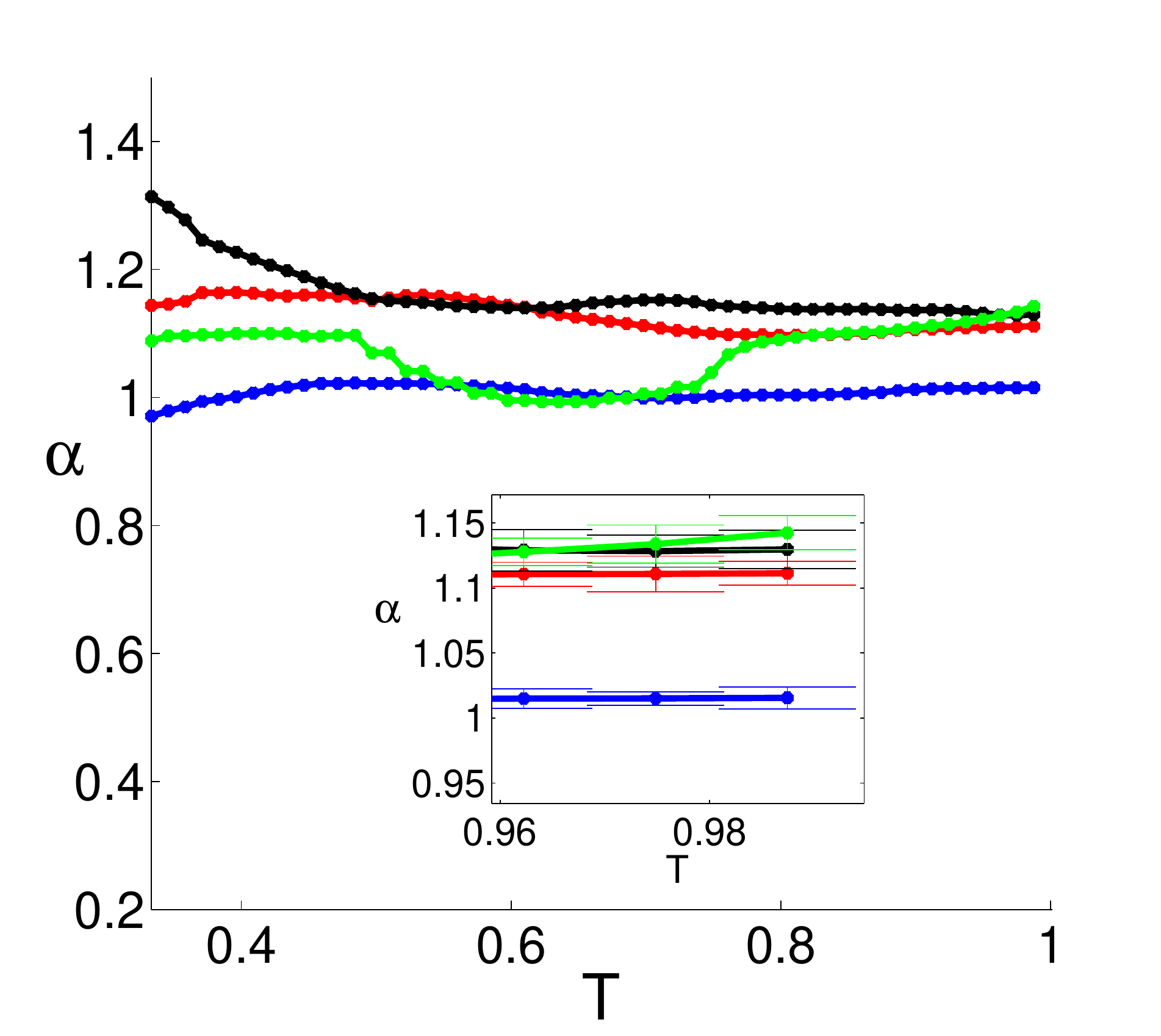}
\caption{\textbf{Convergence of the $\alpha$ exponent as a function of time interval in the SA regime}
Critical exponent $\alpha$ (see~\eqref{eq:xy_decay}) as a function of $T = (t_f - t_0)/t_f-80$, where $t_0$ is the variable initial time and $t_f$ is the final time of the evolution, fixed to $16000$ for $\lambda_x=-\lambda_y=2.5$ (green curve)  and to $t_f=30000$ for $\lambda_x=-\lambda_y=0, \; 1.0,\; 1.7$ (blue, red and black curves respectively).  We observe a saturation of the exponent for longer time intervals, i.e. $T\to 1$, consistent with the fact that the universal dynamical behaviour~\eqref{eq:xy_decay} is expected to appear in the late time dynamics. The \emph{inset} emphasises the behaviour for the last points, with error bars due to the statistical uncertainty.
}
\label{fig:convergenceAlpha}
\end{figure}

%
%
\begin{figure}
\includegraphics[width=0.9\linewidth]{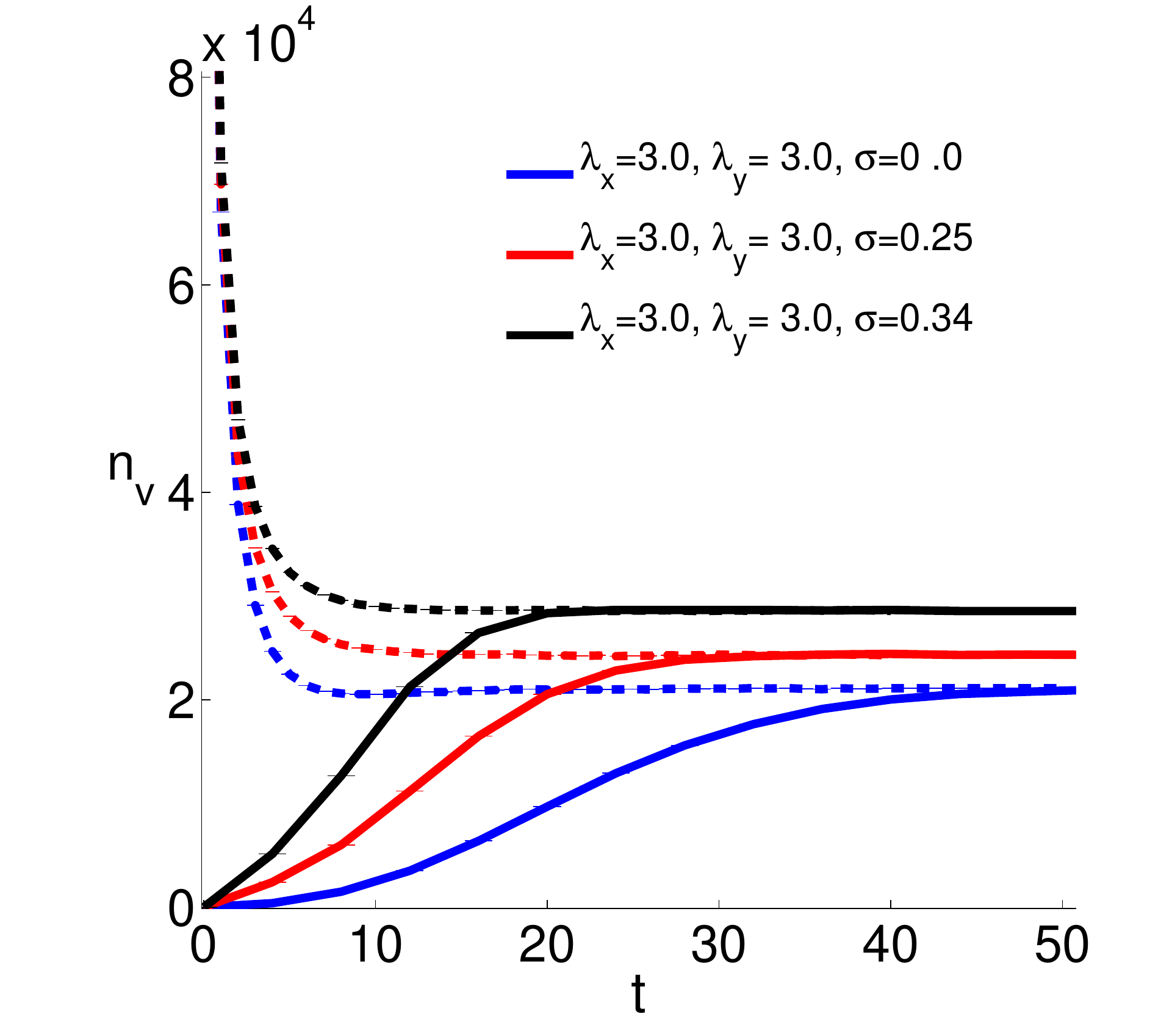}
\caption{\textbf{Steady-state vortex number for the isotropic case.}
Vortex number as a function of time $t=\Delta t\cdot N_{\textrm{time}}$ for $\lambda_x=\lambda_y=3.0$. The dashed lines indicate the time evolution from an initial configuration with a high number of vortices ($n_{\textrm{v}}\sim 2\cdot 10^5$), whereas the continuous lines are from an initial configuration with no vortices. We observe that the steady-state $n_{\textrm{v}}$ does not depend on the initial configuration.
}
\label{fig:convergenceNv}
\end{figure}
%

Now, we examine the convergence  with respect to the size of the lattice. As an example, we study the $\lambda_x=\lambda_y=1.9$ and $\Delta t=0.025$ for $N=1024$ and $N=1200$.
In Fig.~\ref{fig:convergenceLattice} we show that the convergence of the vortex density $\rho$ with respect to the system size $N$ has been reached. 

Next, we study the convergence with respect to the number of stochastic realisations $M$. We choose the case of the XY model with $N=1024$ and $\Delta t=0.025$. Specifically, we look at the numerical exponent $\alpha$ of the vortex equation~\eqref{eq:xy_decay}.
In Fig.~\ref{fig:convergenceRealisations}, we show that the $\alpha$ exponent already converges for $M>100$. However, a detailed and accurate study of the critical dynamics of a stochastic system requires a high number of realisations~\cite{comaron2017dynamical} and, consequently, we consider $M=200$ for all the data presented in the main study.

Finally, we study the convergence of $\alpha$ as a function of the time interval $T$ of the evolution used for the fit for all four sets of parameters used (see Fig.~2 in the main text). 
We consider the dynamical evolution expressed in~\eqref{eq:xy_decay} in a time interval between $[t_0,t_f]$, where the fixed final time $t_f$ takes the value $t_f=16000$ for $\lambda_x=-\lambda_y=2.5$ and $t_f=30000$ for $\lambda_x=-\lambda_y=0, \; 1.0,\; 1.7$. We study the behaviour of the $\alpha$ exponent as a function of the initial time $t_0$ taken to be between $80$ and $2/3\cdot t_f$. Results are displayed in Fig.\ref{fig:convergenceAlpha}, where we show $\alpha$ as a function of $T=(t_f - t_0)/t_f-80$ and consequently $T\in[0,1]$. $\alpha$ converges for $\lambda_x=-\lambda_y=0, \; 1.0,\; 1.7$ configurations when $T\to 1$ as expected since the universal behaviour of~\eqref{eq:xy_decay} should appear in the long time dynamics. 
As we discuss in the main text, we observe that the value of $\alpha$ is somewhat larger for finite $\lambda$ than for $\lambda=0$. This effect can be a consequence of the enhancement of V-AV potential for finite $\lambda$'s. However, the change of the exponent is small with respect to the exponent obtained numerically for the XY model (less than 20$\%$). A deeper analysis of the nature of this difference is out of the scope of the present study.

\section{Steady-state: convergence of the vortex number}
\label{sec:convMag}

\begin{figure}
\includegraphics[width=1.0\linewidth]{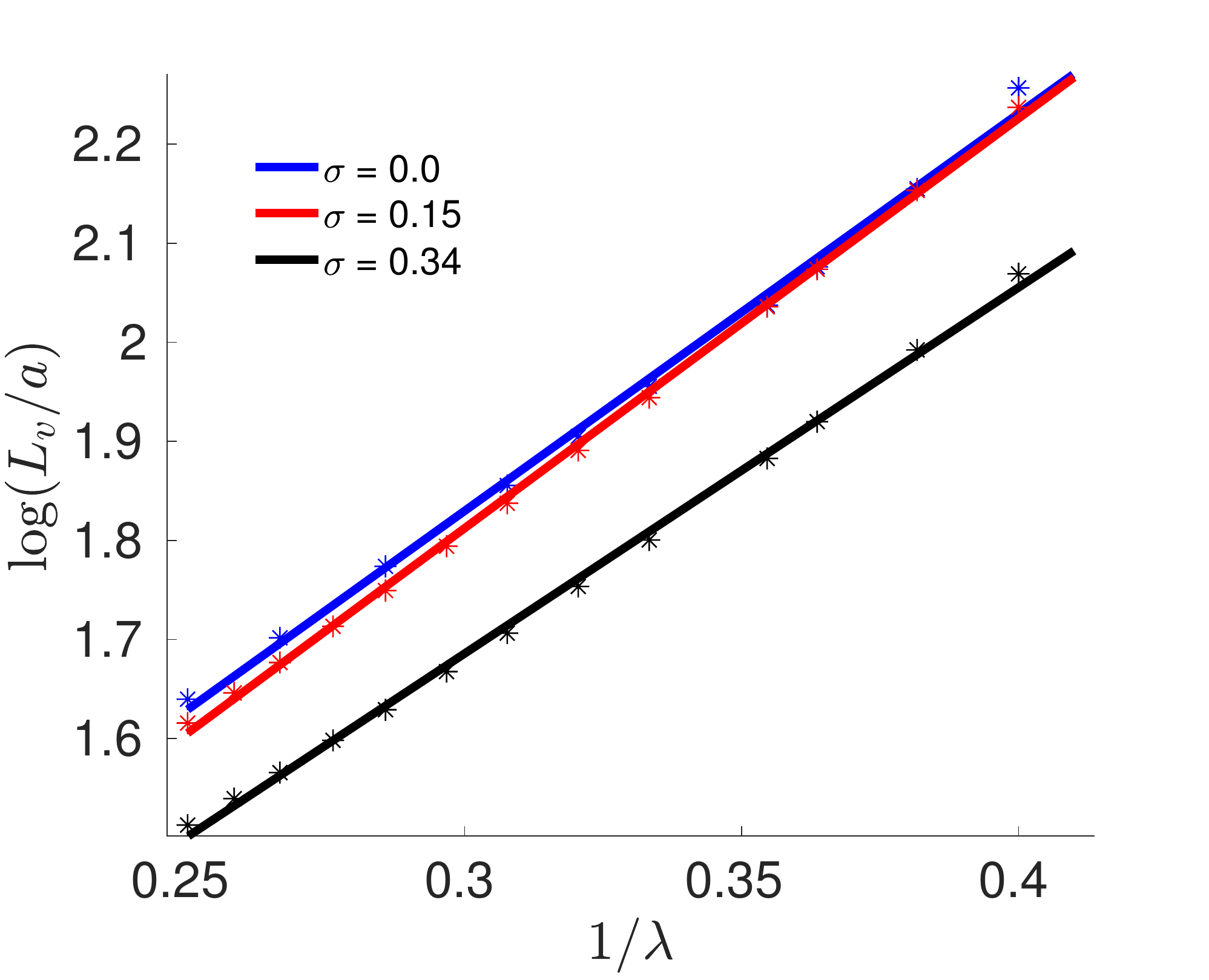}
\caption{\textbf{Characteristic length scale of the vortex-dominated phase in the isotropic system.}
The length scale $L_v$, beyond which the interactions between vortices and anti-vortices become repulsive, as a function of the inverse of the non-linearity $2.5 \leq \lambda\leq 4.0$ and noise levels: $\sigma=0.0, \; 0.15,\; \sigma_q=0.34$.
}
\label{fig:convergenceLv}
\end{figure}
%

As shown in Fig.~1 of the main text, the steady-state of the cKPZ equation in the SA regime shows i) a low noise phase with finite magnetization $M$ (where $M^2=\langle  (\sum\limits_i \frac{1}{N^2}\cos\theta_i)^2 +
(\sum\limits_i \frac{1}{N^2}\sin\theta_i )^2 \rangle$ and $\langle \cdot\cdot\cdot\rangle$ denotes averaging over realisations) and low density of paired vortices ii) a high noise phase with zero $M$ and high density of entropic free vortices.
However, the system does not show such phases in the isotropic regime. In contrast, there is a vortex dominated phase even at low or zero noise level. The density of vortices appears to be independent of the noise when this goes to zero, as has been recently predicted~\cite{wachtel2016electrodynamic,sieberer2016lattice}, provided that the average distance between vortices is shorter than the system size. This occurs when $\lambda>2.5$ at the zero noise level for our $1024\times1024$ square lattice.

We obtain the steady state of the system by numerically integrating the  cKPZ equation on a lattice \eqref{eq:cKPZLattice} for a number of time iterations $N_{\textrm{time}}$ (which depends strongly on the noise level and the non-linearity of the system) from an initial configuration with no vortices, where the initial phase at each point takes a random value in the interval $[-\pi/4,\pi/4]$.

Particularly, for cases in the SA regime, $N_{\textrm{time}}\geq 320000$ when $\sigma<\sigma_c$, since the convergence of the magnetisation requires long times. However, for the isotropic case and the SA cases in the disordered phase ($\sigma>\sigma_c$), we obtain convergence to the steady-state for $N_{\textrm{time}}\geq 8000$, since the system reaches $M=0$ and non-zero vortex density in this scenario, which converges quickly.

Next, we check that the high density of vortices observed at low noise level for large $\lambda_x=\lambda_y$  truly corresponds to the steady-state of the system and is not an artefact of the initial conditions.
Specifically, we consider the $\lambda_x=\lambda_y=3.0$ configuration, shown in Fig.~1 of the main text, and compare the vortex number as a function of time for three different low noise levels ($\sigma=0.0, \; 0.25$ and $\sigma=0.34$) with two very different initial conditions: 
i) a configuration with a high number of vortices ($n_{\textrm{v}}\sim 2\cdot 10^5$), where the initial phase at each point takes a random value in the interval $[-\pi,\pi]$ and ii) a configuration with no vortices, where the initial phase at each point takes a random value in the interval $[-\pi/4,\pi/4]$. The results are shown in Fig.~\ref{fig:convergenceNv}. We observe that both initial conditions result in exactly the same steady-state number of vortices. This vortex number depends on the noise level (see also Fig.~1 in the main text) since a non-zero noise introduces entropic vortices in addition to the ones appearing purely due to the repulsive interactions. The non-zero number of vortices for $\sigma=0$ is a clear indication of the new vortex dominated non-equilibrium phase.

Finally, we look at $L_v$, given by Eq.~(3) in the main text, which sets the distance beyond which the interactions between vortices and anti-vortices become repulsive. In Fig.~\ref{fig:convergenceLv} we show the results fitted with $L_v/a = A e^{B/\lambda}$ for $\lambda_x=\lambda_y=3.0$ and three different noise levels: $\sigma=0.0$, $0.15$ and $\sigma=\sigma_q=0.34$, which is the noise level used for the infinite rapid quench in the main text. The results are: i) $\log(A) = 0.728 \pm 0.022$, $B=3.672 \pm 0.092$ ($\sigma=0.0$) ii) $\log(A) = 0.699 \pm 0.044$, $B=3.72 \pm 0.18$ ($\sigma=0.15$) and iii) $\log(A) = 0.734 \pm 0.058$, $B=3.173 \pm 0.24$ ($\sigma=\sigma_q$, see inset of Fig. 3 in the main text).
We  observe that noise levels beyond $\sigma>0.15$ introduce additional entropic vortices that seem to cause a small reduction in $L_v$.

\section{Fitting functional form to the vortex decay}
\label{sec:different}

\begin{figure}
\includegraphics[width=1.05\linewidth]{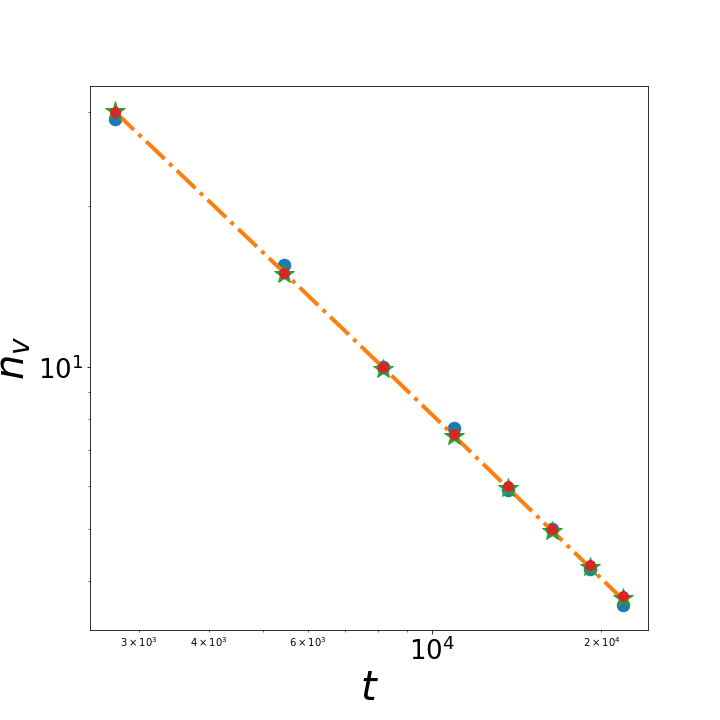}
\caption{\textbf{Comparison of different fitting functions for the vortex decay.}
Number of vortices as a function of time  from numerical simulation of the  $\lambda_x=-\lambda_y=1.7$ case (blue dots), together with $(\log(t)/t)^\alpha$ fit (dashed-dotted line) and $(1/t)^\gamma$ fits (green 'star' dots  and red dots for $\gamma=1.006$ and $\gamma=1.00$ cases respectively). As mentioned in the text, the decay $(\log(t)/t)^\alpha$ gives the best fit, with $\alpha\approx 1.13$.
}
\label{fig:otherPowerLaw}
\end{figure}
%

\begin{figure}
\includegraphics[width=1.0\linewidth]{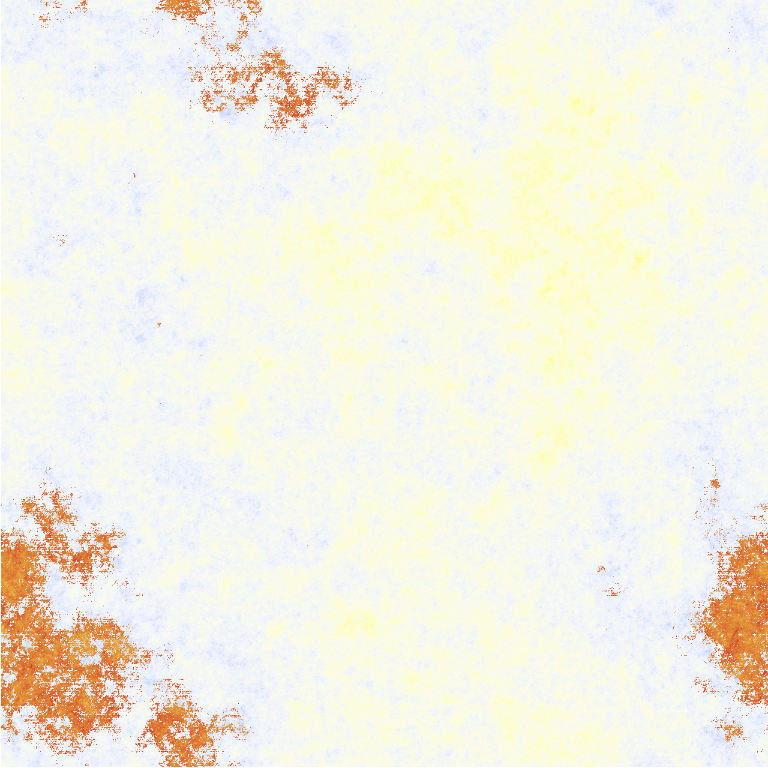}
\includegraphics[width=1.0\linewidth]{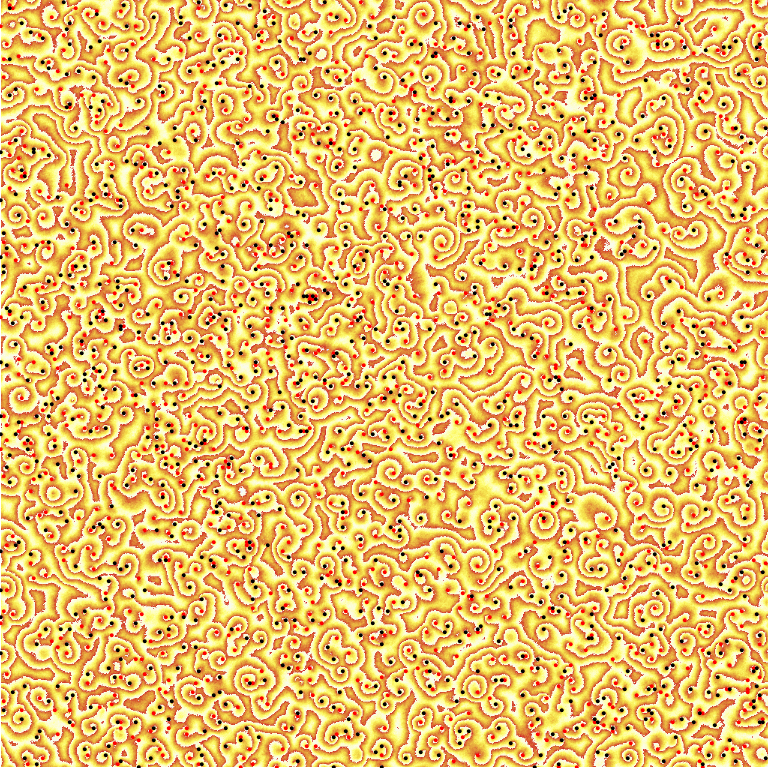}
\caption{\textbf{$\theta(\ve{r})$ at late times with marked vortices.} 
2D map with position of vortices (black dots) and antivortices (red dots) on top of the phase configuration for a single realisation of i) a cKPZ system in the SA regime with $\lambda_x=-\lambda_y=1.7$ (top panel), note that there are zero V-AV pairs in that particular configuration and ii) a cKPZ system in isotropic regime with $\lambda_x=\lambda_y=1.5$ (bottom panel).
}
\label{fig:phases}
\end{figure}
%

As we discussed in section \emph{Diffusive decay of the vortex density} of the main paper and Sec.~\ref{sec:convergence} of the present document, the vortex decay in the SA regime follows a diffusive law with a logarithmic correction $n_v\sim (\log(t) /t)^\alpha $ and an exponent $\alpha$ close to one. 
As in the equilibrium XY-model, this type of functional form fits the data better than the $n_v\sim (1/t)^{\gamma}$ decay.
Specifically, for the KPZ system with $\lambda_x = -\lambda_y= 1.7$, we find that $\alpha\approx 1.13$ with a coefficient of determination $R^2 = 0.99889$, whereas fitting  $n_v\sim (1/t)^{\gamma} $ gives $\gamma \approx 1.006$ with $R^2 = 0.99863$ and $R^2 = 0.99849$ when $\gamma = 1.00$
\footnote{As a reference, for the equilibrium XY-model case, i.e. $\lambda_x = \lambda_y= 0$, we find  an exponent $\alpha\approx 1.01$ with a fit with coefficient of determination $R^2 = 0.99805$, whereas a fit $n_v\sim (1/t)^{\gamma} $ gives $\gamma \approx 0.89$ with $R^2 = 0.99787$.
}.
The three different fits are displayed in Fig.~\ref{fig:otherPowerLaw}. Although the form with logarithmic corrections fits our data somewhat better then the pure power-law decay, the difference is not large and so further investigations, beyond the scope of this work, would be needed to provide a definite answer about the faith of logarithmic corrections in the KPZ phase ordering.

\section{$\theta(\ve{r})$  spatial profile and vortices}

As mentioned in the main paper, the cKPZ system in the SA regime seem to show a diffusive decay of the number of vortices  with a characteristic logarithmic correction, due to the attractive V-AV interaction. Consequently, at long times the system eventually arrives in a state with  a relatively  smooth  $\theta(\ve{r})$  and no vortices due to the V-AV annihilation. As an example, in the top panel of Fig.~\ref{fig:phases}, we show a snapshot of the $\theta(\ve{r})$  profile after a long time dynamics  for a system in the SA regime, with $\lambda_x=-\lambda_y=1.7$.  This particular realisation has no vortices. A different realisation for the same parameters shown in the left panel of Fig 2 of the main paper has three pairs of vortices. Averaging over realisations leads to finite but small number of vortices in this regime.

On the other hand, a cKPZ system in the isotropic regime  converges at the vortex-dominated phase due to the repulsive interactions between the Vs and AVs at large distances.
As an example, in Fig.~2 of the main paper we showed  a 2D map of $\theta(\ve{r})$ with marked positions of vortices for $\lambda_x=\lambda_y=1.5$ zoomed to the quarter of the whole system for better visibility of $\theta(\ve{r})$ structure. In this section in Fig.~\ref{fig:phases} we present the whole system 2D map for this particular realisation to highlight the highly disordered $\theta(\ve{r})$ and a large number of vortices.
Additionally, we can observe the spiral configuration of $\theta(\ve{r})$, which is characteristic of a cKPZ system in the WA regime \cite{sieberer2018topological}.

%


\end{document}